\documentclass[onecolumn,preprint3]{aastex63}
\usepackage{lineno}
\usepackage{bbding}
\usepackage{amssymb,amsmath,bm}
\usepackage{rotating}
\usepackage{hyperref}
\usepackage{bm}
\hypersetup{
  colorlinks      = {true},
  linkcolor       = {blue},
  citecolor       = {blue},
  urlcolor        = {blue},
}

\shorttitle{}
\shortauthors{}
\begin{document}
\title{Model-independent test for the cosmic distance duality relation with Pantheon and eBOSS DR16 quasar sample }

\author{Bing Xu}
\affiliation{School of Electrical and Electronic Engineering, Anhui Science and Technology University, Bengbu, Anhui 233030, China}

\author{Zhenzhen Wang}
\affiliation{Department of Physics, Anhui Normal University, Wuhu, Anhui 241000, China}

\author{Kaituo Zhang}
\affiliation{Department of Physics, Anhui Normal University, Wuhu, Anhui 241000, China}
\email{ktzhang@ahnu.edu.cn}

\author{Qihong Huang}
\affiliation{School of Physics and Electronic Science, Zunyi Normal University, Zunyi 563006, Guizhou, China}

\author{Jianjian Zhang}
\affiliation{School of Electrical and Electronic Engineering, Anhui Science and Technology University, Bengbu, Anhui 233030, China}

\begin{abstract}

In this paper, we carry out a new model-independent cosmological test for the cosmic distance duality relation~(CDDR) by combining the latest five baryon acoustic oscillations (BAO) measurements and the Pantheon type Ia supernova (SNIa) sample. Particularly, the BAO measurement from extended Baryon Oscillation Spectroscopic Survey~(eBOSS) data release~(DR) 16 quasar sample at effective redshift $z=1.48$ is used, and two methods, i.e. a compressed form of Pantheon sample and the Artificial Neural Network~(ANN) combined with the binning SNIa method, are applied to overcome the redshift-matching problem. Our results suggest that the CDDR is compatible with the observations, and the high-redshift BAO and SNIa data can effectively strengthen the constraints on the violation parameters of CDDR with the confidence interval decreasing by more than 20 percent.  In addition, we find that the compressed form of observational data can provide a more rigorous constraint on the CDDR, and thus can be generalized to the applications of other actual observational data with limited sample size in the test for CDDR.

\end{abstract}


\section{Introduction}\label{S1}

Based on three fundamental hypotheses that the spacetime is described by a metric theory of gravity, photons always travel along null geodesics and their number is conserved, \citet{Etherington93,Etherington07} proved a famous cosmic distance-duality relation (CDDR), which connects the luminosity distance (LD) $D_{\rm L}$ and the angular diameter distance (ADD) $D_{\rm A}$ at the same redshift $z$ through the following identity

\begin{equation}
\frac{D_{\rm{L}}(z)}{D_{\rm{A}}(z)}(1+z)^{-2}=1\,.
\end{equation}
This relation is independent of the Einstein field equations and the nature of matter, and has been widely used in astronomical observations and modern cosmology as a fundamental relation. For example, the CDDR has been used to test the geometrical shape of galaxy clusters~\citep{Holanda11}, the temperature profile and the gas mass density profile of galaxy clusters~\citep{Cao11,Cao16}. However, a violation of one of the hypotheses leading to the CDDR might be possible, which may be considered as a signal of exotic physics~\citep{Bassett04a,Bassett04b}. Therefore, it is necessary to test the reliability of the CDDR accurately before applying to various astronomical theories.

A straightforward approach to test the validity of CDDR is to constrain the parameterization function $\eta(z)\equiv{D_{\rm{L}}(z)}/{D_{\rm{A}}(z)}(1+z)^{-2}$ with the LD and ADD of some objects at the same redshift. Here the function $\eta(z)$  represents the possible violation of the standard CDDR, and can be parameterized in distinct forms, such as $\eta(z,\eta_1)=1+\eta_1 z$, $\eta(z,\eta_2)=1+\eta_2 z/(1+z)$, and $\eta(z,\eta_3)=1+\eta_3 \mathrm{ln}(1+z)$, where the parameter $\eta_i$ can be constrained by observational data. Following this idea, a lot of works have been devoted to test the validity of CDDR by combining LD data inferred from the observations of type Ia supernovae~(SNIa), HII galaxies, or gamma-ray burst with ADD data determined from different observations such as the X-ray plus Sunyaev-Zel'dovich~(SZ) effect of galaxy clusters, and the gas mass fraction measurement in a galaxy cluster~\citep{Uzan04,Bernardis06,Holanda10,Holanda11,Holanda12,Li11,Nair11,Meng12,Yang13,Santos15,Hu18,Silva20,Bora21}, the angular size of ultra-compact radio sources~\citep{Li18,Liu21}, and strong gravitational lensing~\citep{Liao16,Holanda16,Holanda17,Ruan18,Lima21,Qin21}. These works do not suggest that the CDDR deviates significantly from the real universe. In addition, since the baryon acoustic oscillations (BAO) measurement, which is a very precise experiment and plays an important role in modern cosmology, can provide more accurate ADD data than the observations described above, it is also used to test the CDDR and has been proved to be a very powerful tool to test the relation~\citep{Wu15}. Recently, some works have attempted to test the CDDR with LD data from different SNIa observations and the ADD data derived from different BAO measurements. For example, combining the Union2.1 SNIa data~\citep{Suzuki21} with five BAO ADD data points from the WiggleZ Dark Energy Survey~\citep{Blake12}, the Sloan Digital Sky Survey~(SDSS) Data Release 7~(DR7)~\citep{Xu13} and DR11~\citep{Samushia14}, \citet{Wu15} tested the validity of CDDR and found that there was no violation of CDDR.  A similar result was obtained by~\cite{Lin18}, in which the authors tested the validity of CDDR by using the same BAO data points plus the ADD data from galaxy clusters and the latest SNIa sample, i.e. the Pantheon sample from the Pan-STARRS1 Medium Deep Survey which is the largest SNIa sample released to date and consists of 1048 SNIa data covering the redshift range of $0.01<z<2.3$~\citep{Scolnic18}. \citet{Xu20} updated the constraints with the Baryon Oscillation Spectroscopic Survey~(BOSS) DR12 data and the Pantheon sample, and they obtained consistent results. However, it should be noted that although the redshift of the newest Pantheon SNIa sample  is up to $z\sim2.3$, these tests still suffer from low redshift range, i.e. $z<1$, due to the lack of  observations at higher redshift. Most recently, eBOSS collaboration provided a precise BAO measurement from the final quasar sample of BOSS DR16 at effective redshift $z=1.48$~\citep{Neveux20,Alam21}. In this paper, we therefore plan to check the validity of CDDR by combining the newest BAO measurements with the Pantheon sample. Using these data, the tests of CDDR with the ADD data derived from BAO measurement can reach high redshift range $z \sim 1.5$.

Here, it should be pointed out that since any deviation from the CDDR can contribute to the non-conservation of the photon number~\citep{Ellis07}, exploring the CDDR is equivalent to testing the cosmic opacity of the universe. Thus, the parameter of the CDDR can also be constrained by the parametrization function of optical depth $\tau(z)$ rather than $\eta(z)$ with their relation being $\eta(z)=e^{\tau(z)/2}$~\citep{Lima11}. And many works have been made to perform the constraints on the cosmic opacity by using various astronomical observations,  and the results show that there is no obvious evidence of an opaque universe~\citep{More09, Nair12, Chen12, Li13, Holanda13, Liao13, Liao15, Wang17,Ma19, Qi19b, Wei19, Zhou19, Liu20, Fu20, Geng20, Xu21, He22}. However, it is worth mentioning that combining the simulated gravitation waves from DECi-hertz Interferometer Gravitational-wave Observatory~(DECIGO) and the Einstein Telescope~(ET) with the observations of SNIa, HII galaxies and  monochromatic X-ray and ultraviolet (UV) luminosity of quasars, the authors in the references~\citep{Liu20,Geng20} have tested the cosmic opacity out to high redshifts and found that the constraint results is slightly sensitive to the parameterization of $\tau(z)$. In this paper, considering that the expressions of $\eta(z)$ have several advantages such as a manageable one-dimensional phase space and a good sensitivity to observational data~\citep{Holanda11}, we only use the parameterizations of $\eta(z)$ as listed in the previous text to check the validity of CDDR.

A common issue we will encounter when performing the tests is that measurements of LD from SNIa data and ADD from BAO data are not available at the same redshift. In response to this issue, several methods have been proposed in the literatures. \citet{Holanda10} and \citet{Li11} derived the LD by using the SNIa data whose redshift is closest to the cluster's within the range $\Delta z=|z-z_{\rm SNIa}|<0.005$. \citet{Meng12} extended the method by binning all SNIa data available in the range $\Delta z$. \citet{Cardone12} applied a local regression technique to the SNIa data at redshift windows of interest with adjustable bandwidth. These methods fail at $z=1.48$ because the current SNIa data have very few distributions at high redshifts $z>1$. The values of LD at the redshift of BAO measurements can also obtain by using the Gaussian Processes~(GPs)~\citep{Nair15,Rana17,Mukherjee21}. However, when the number of observed events is not enough, GPs are not reliable  for the reconstruction of SNIa~\citep{ZhouandLi19} and lead to great uncertainty~\citep{Lin18}. In addition, \citet{Ma18} applied a Bayesian statistical method to calculate the SNIa luminosity distance moduli at the redshifts of ADD data from BAO measurements. Although this method can effectively reduce the uncertainty and consider the correlation between data points, it needs to assume auxiliary cosmological information to obtain the values of LD and ADD. In this paper, we use two methods to overcome the redshift-matching problem. Firstly, we propose a new method to overcome the issue. In this method, a compressed form of the Pantheon SNIa sample is provided by using a piecewise linear function of $\mathrm{ln}(z)$ proposed by~\citet{Betoule14}, which is shown that it still remains accurate for cosmological constraints. By this methods, we can derive the LD values at the redshift of BAO data points self-consistently from the binned apparent magnitude values based on the linear function. Most recently, an Artificial Neural Network~(ANN) is used to reconstruct the Hubble diagrams from the observed HII galaxy and radio quasar samples and then test the CDDR~\citep{Liu21}. The main purpose of using an ANN is to construct an approximate function that associates input data with output data. This machine learning reconstruction method is data-driven and makes no assumptions about the data, which suggests that it is completely model-independent, and have shown outstanding performance in solving cosmological problems in both accuracy and efficiency, such as analyzing gravitational waves~\citep{Li20,George18}, estimating cosmological parameters~\citep{Fluri19,Wang20b,Wang21}, and studying the evolution of dark energy models~\citep{Escamilla-Rivera20}. Specially, it has been shown to be able to reconstruct a safer function than GPs when there are fewer data points~\citep{Wang20a}.  In order to see whether the conclusions are changed by a different method and  improve the robustness of the test, we will then apply the ANN method to reconstruct the apparent magnitude-redshift relation $m_{\rm B}(z)$ which is used to test the CDDR by combining the newest BAO measurements. Furthermore, in our analysis, the nuisance parameters used in the construction of likelihood function, such as the absolute B-band magnitude $M_{\rm B}$ and the fiducial value of sound horizon $r_{\rm d,f}$, are marginalized  analytically with flat priors to avoid the tests that rely on any auxiliary cosmological information.

This paper is organized as follows: we introduce the data and methodology used in our analysis in Section 2. In section 3, we apply the method to test the CDDR and yield results. Conclusions and discussions are presented in Section 4.


\section{Data and Methodology}\label{S2}

In this work, we aim to use ADD derived from the newest BAO measurements and LD derived from the latest Pantheon sample to test the CDDR by constraining the function $\eta(z)$ parameterized with $\eta_i$. So we now introduce briefly the BAO data used in our analysis. The BAO refers to an overdensity or clustering of baryonic matter at certain length scales due to the oscillations of acoustic waves which propagated in the early universe. It happens on some typical scales and can provide a standard ruler for length scale in cosmology to explore the expansion history of the universe. The length of this standard ruler ($\sim$150 Mpc in today's universe) corresponds to the distance that a sound wave propagating from a point source at the end of inflation would have traveled before decoupling. Most recently, the eBOSS Collaboration have released their final BAO observations and summarized fourteen measurements of $D_{\rm V}(z)/r_{\rm d}$, $D_{\rm M}(z)/r_{\rm d}$ and $D_{\rm H}(z)/r_{\rm d}$, covering the redshift range $0.15\leq z \leq2.33$~(see Tabel III in Ref.~\citep{Alam21}). Here, $r_{\rm d}$ is the sound horizon, $D_{\rm V}(z)$, $D_{\rm M}(z)$ and $D_{\rm H}$ is the spherically-averaged distance, the comoving ADD and the Hubble distance, respectively. These measurements are obtained from the final observations of clustering using galaxies, quasars, and Ly$\alpha$ forests from the completed SDSS lineage of experiments in large-scale structure, composing of data from SDSS, SDSS-II, BOSS, and eBOSS, and thus allow us to perform a comprehensive assessment of the cosmological model and parameter. In this paper, given that the largest redshift of Pantheon sample $z<2.3$~\citep{Scolnic18} and the ADD is associate with the comoving ADD through the relation $D_{\rm M} = (1 + z)D_{\rm A}$, we will only use the five measurements of $D_{\rm M}(z)/r_{\rm d}$ from the SDSS-III BOSS DR12 galaxy sample~\citep{Alam17}, the SDSS-IV  eBOSS DR16 LRG sample~\citep{Marin20, Bautista21}, the SDSS-IV eBOSS DR16 ELG sample~\citep{Tamone20, Mattia21} and the SDSS-IV eBOSS DR16 quasar  sample~\citep{Neveux20,Hou21}, which are  summarized  in  Table~\ref{Tab1}. Here, it is necessary to point out that although the BAO measurements are widely used in cosmological analysis, the so-called fitting problem still remains a challenge for BAO peak location as a standard ruler~\citep{Ellis87}. In particular, the environmental dependence of the BAO location has recently been detected by~\citet{Roukema15,Roukema16}. Moreover, \citet{Ding15} and \citet{Zheng16} pointed out a noticeable systematic difference between H(z) measurements based on BAO and those obtained with differential aging techniques. Since these problems are related to the calibration of $r_{\rm d}$, we will therefore introduce a fiducial value of sound horizon $r_{\rm d,f}$ as a nuisance parameter  to calculate the values of ADD from the measurements of BAO and marginalize its influence with a flat prior in the analysis.

\begin{table}
  \caption{Summary of the dimensionless $D_{\rm M}(z)/r_{\rm d}$ measurements from different BAO samples.}\label{Tab1}
   \begin{center}
\addtolength{\leftskip} {-1.5cm}
    \begin{tabular}{cccc}
    \hline
    \hline
       $z_{\rm eff}$  & $\mathrm{value}$ & $\mathrm{Survey}$ & $\mathrm{Reference}$  \\
    \hline
     $0.38$ & $10.23\pm0.17$ & $\mathrm{BOSS \, Galaxy}$ & \citet{Alam17} \\
    \hline
     $0.51$ & $13.36\pm0.21$ & $\mathrm{BOSS \, Galaxy}$  & \citet{Alam17} \\
    \hline
     $0.70$ & $17.86\pm0.33$ & $\mathrm{eBOSS \, LRG}$ & \citet{Marin20} \\
    \hline
     $0.85$ & $19.5\pm1.0$ & $\mathrm{eBOSS \, ELG}$ & \citet{Tamone20} \\
    \hline
     $1.48$ & $30.69\pm0.80$ & $\mathrm{eBOSS \, Quasar}$ & \citet{Neveux20} \\
    \hline
    \end{tabular}
\end{center}
\end{table}

Now we turn our attention on the Pantheon compilation released by the Pan-STARRS1 Medium Deep Survey~\citep{Scolnic18}, which consists of 1048 SNIa data covering the redshift range $0.01< z < 2.3$. And the observed distance modulus of each SNIa in this compilation is given by

\begin{equation}
\mu_{\rm obs}(z)=m^*_{\rm B}-(M_{\rm B}-\alpha X_1+\beta\mathcal{C})\,.
\end{equation}
Here, $m^*_{\rm B}$ is the observed peak magnitude in rest frame B-band, $X_1$ is the time stretching of the light-curve, $\mathcal{C}$ is the SNIa color at maximum brightness, $M_{\rm B}$ is the absolute magnitude, $\alpha$ and $\beta$ are two nuisance parameters, which should be fitted simultaneously with the cosmological parameters. In order to avoid the dependence of $\alpha$ and $\beta$ on the cosmological model, \citet{Kessler17} proposed a new method called BEAMS with Bias Corrections~(BBC) to calibrate the SNIa, and the corrected apparent magnitude $m^\ast_{\rm B,corr} = m^\ast_{\rm B}+\alpha X_1 - \beta C +\Delta_{\rm B}$ for all the SNIa is reported in Ref.~\citep{Scolnic18} with $\Delta_{\rm B}$ being the correction term. And the observed distance modulus is rewritten as

\begin{equation}
\mu_{\rm obs}=m^*_{\rm B,corr}-M_{\rm B}\,.
\end{equation}

To test the validity of the CDDR with the BAO measurements and Pantheon SNIa sample, we must obtain the values of LD from SNIa data at the redshifts of BAO data. For this purpose,  we use two methods to derive the apparent magnitude values at the redshifts of BAO measurements, in order to achieve the CDDR testing and obtain convincing results. In the first way, we provide the cosmological information of the Pantheon SNIa sample in a compressed form with the method proposed by~\citet{Betoule14}. Here, instead of binning the distance modulus used in~\citep{Betoule14}, the corrected apparent magnitude is approximated by a piecewise linear function of $\mathrm{ln}(z)$ by defining on each segment $z_b \leq z \leq z_{b+1}$ as:

\begin{equation}\label{IF}
\overline{m}_{\rm B}(z)=(1-\alpha)m_{{\rm B},b}+\alpha m_{{\rm B},b+1}\,.
\end{equation}
Here $\alpha= \mathrm{ln}(z/z_b )/\mathrm{ln}(z_{b+1}/z_b )$, $m_{{\rm B},b}$ is the apparent magnitude at $z_b$. In consideration of the smallest and largest redshift of the Pantheon sample, we use 36 ln-spaced control points $z_b$ in the redshift range $0.01<z<2.3$. Then using the above interpolation function to fit the Pantheon data, we can obtain the compressed form by minimizing the $\chi^2$ function,

\begin{equation}
\chi^2 = \left[\mathbf{m}^*_{\rm B}-\mathbf{\overline{m}}_{\rm B}\right]^{\rm T} \cdot  \mathbf{Cov}^{-1} \cdot \left[\mathbf{m}^*_{\rm B}-\mathbf{\overline{m}}_{\rm B}\right].
\end{equation}
Here, $\mathbf{Cov}=\mathbf{D}_{\rm stat}+\mathbf{C}_{\rm sys}$ is the covariance matrix, where the statistical matrix $\mathbf{D}_{\rm stat}$ has only diagonal elements and $\mathbf{C}_{\rm sys}$ is the systematic covariance.

\begin{figure}
\centering
\includegraphics[width=0.5\textwidth, height=0.334\textwidth]{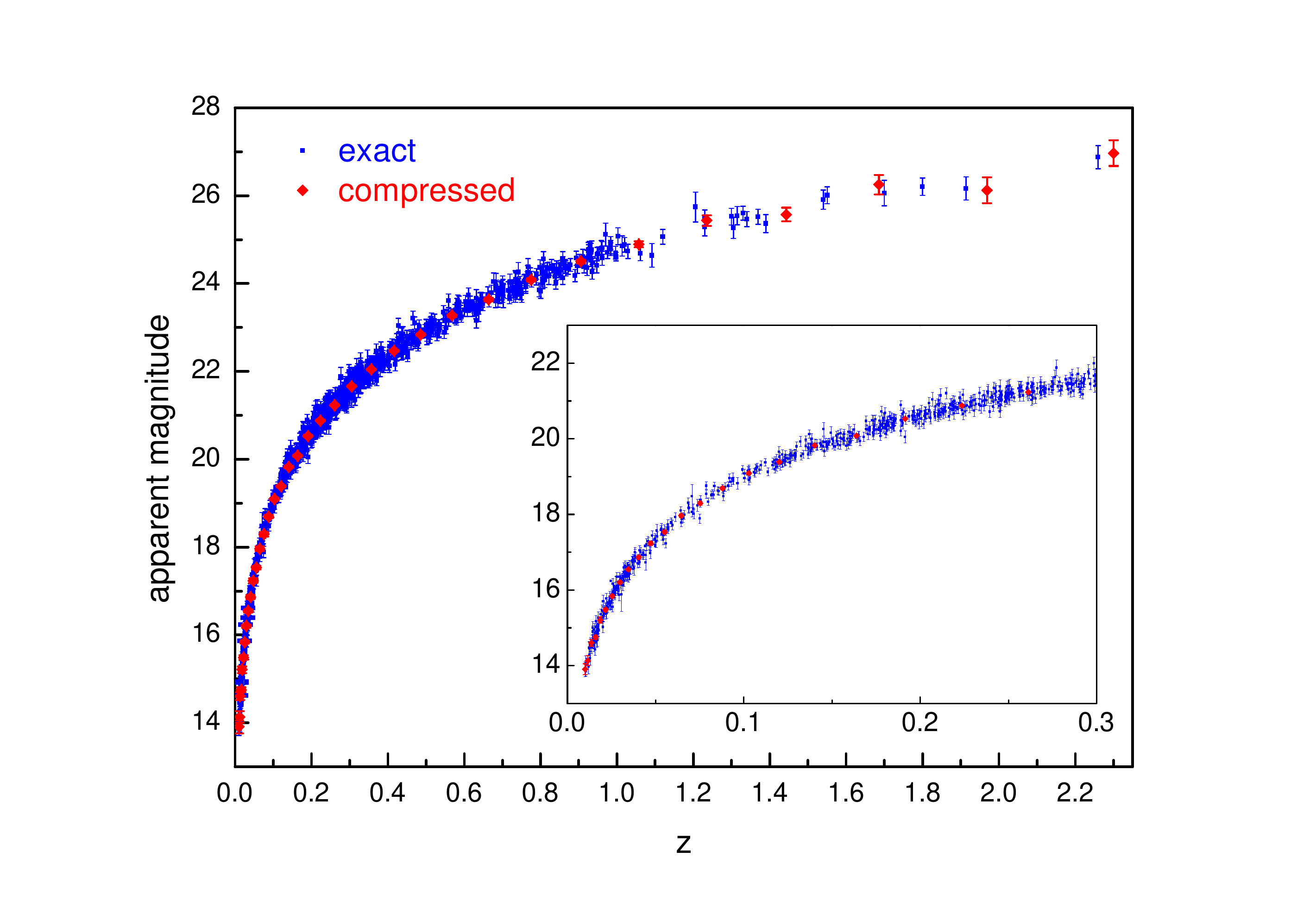}
\includegraphics[width=0.4\textwidth, height=0.3\textwidth]{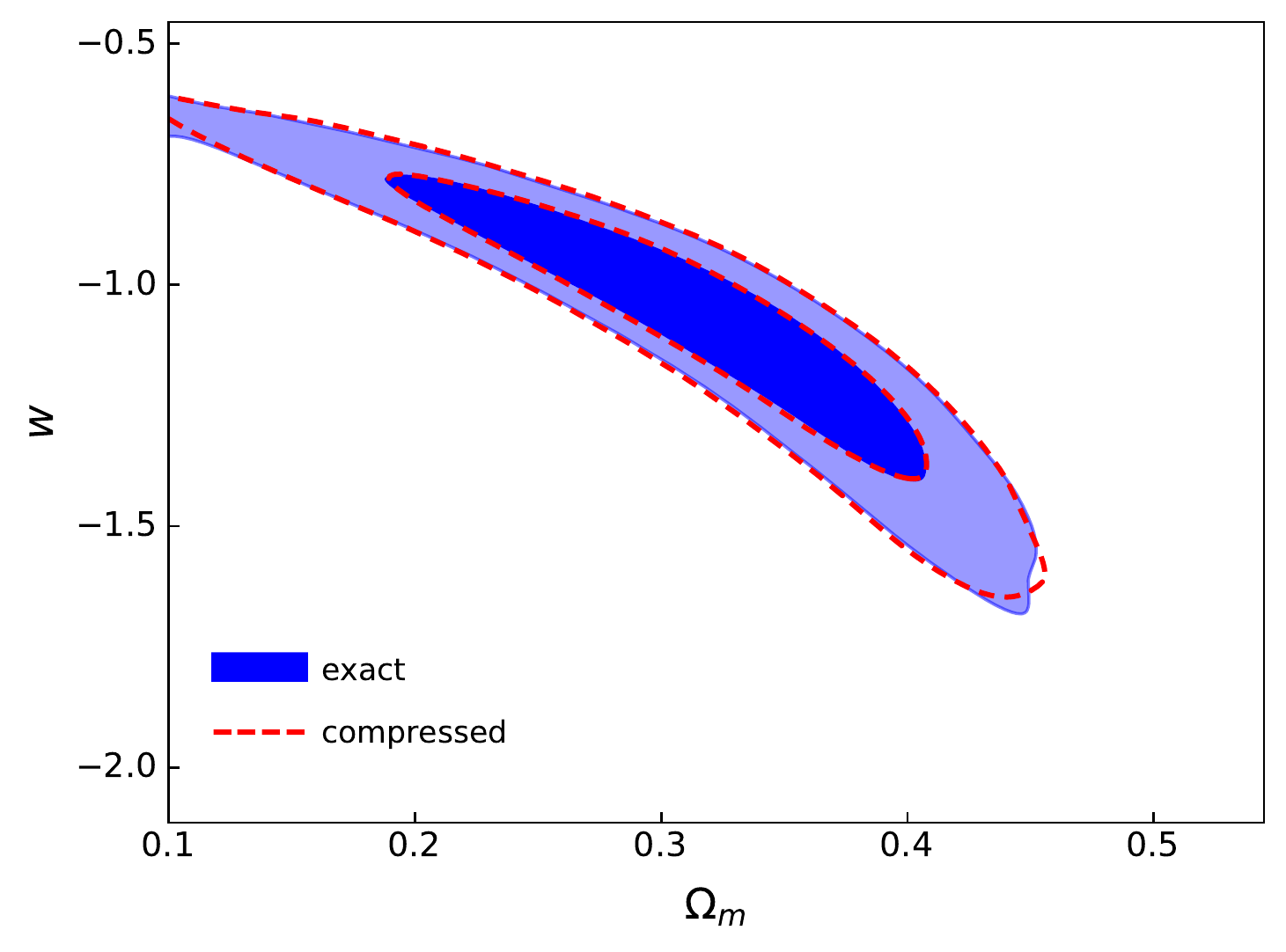}
\caption{ {\bf Left: }The exact measurements of apparent magnitude in Pantheon sample~(blue) and its compressed form~(red). {\bf Right:} Comparison of the cosmological constraints obtained from the exact Pantheon sample likelihood~(filled blue contour) with compressed version~(dashed red contour).}
\label{Fig1}
\end{figure}

In order to constrain the 36 parameter spaces of $m_{\rm B,b}$, we have modified the publicly available Markov Chain Monte Carlo (MCMC) package CosmoMC~\citep{Lewis02} to conduct the calculation. We list the binned apparent magnitude in Table~\ref{Tab2} with their covariance matrix given in Table~\ref{TabA} in the section of Appendix. For a visual comparison, we also plot the exact measurements of apparent magnitude in Pantheon sample and its compressed form in the left of Figure~\ref{Fig1}. One can see that the compressed form of Pantheon sample can provide a good approximation of the relationship between the apparent magnitude and redshift revealed by the full exact measurements. Furthermore, as an illustration, a comparison of the cosmological constraints obtained from the approximate and full version of the Pantheon likelihood for the $w$CDM model is shown on the right side of Figure~\ref{Fig1}. It is clear that the differences of the constraint results from the approximate and full version are very small. This means that the compressed form still remains accurate for cosmological constraints.

Using the compressed form listed in Table~\ref{Tab2} and the interpolation function given in Equation~\ref{IF}, we can derive the values of $\overline{m}_{\rm B}$ from the compressed form of Pantheon sample at the redshift of BAO measurements. Instead of interpolating directly from the two nearest observations, the values of $\overline{m}_{\rm B}$  obtained in our analysis take into account the influence of all observations in two adjacent redshift intervals at a chosen redshift. This can effectively solve the negative impacts caused by the small sample size, such as great uncertainty. We refer to this method as `I' and summarize the interpolated values of $\overline{m}_{\rm B}$ at the redshifts of BAO data in Table~\ref{Tab3}.

\begin{table}
\caption{Binned apparent magnitude fitted to the Pantheon sample.}\label{Tab2}
\begin{center}
\addtolength{\leftskip} {-3.5cm}
 \begin{tabular}{cc|cc|cc|cc}
    \hline
    \hline
       $z_{b}$  & $m_{{\rm B},b}$ & $z_b$ & $m_{{\rm B},b}$ & $z_b$ & $m_{{\rm B},b}$ & $z_b$ & $m_{{\rm B},b}$ \\
    \hline
     0.010 & $ 13.909\pm0.143 $ & $ 0.041 $ & $ 16.862\pm0.046 $ & $ 0.164$ & $ 20.078\pm0.026 $ & $ 0.664 $ & $ 23.633\pm0.042 $\\

     0.012 & $ 14.131\pm0.132 $ & $ 0.047 $& $ 17.234\pm0.050 $ & $0.191$ & $ 20.530\pm0.022 $ & $ 0.775 $ & $ 24.082\pm0.037 $\\

         0.014 & $ 14.604\pm0.088 $ & $  0.055$ & $ 17.536\pm0.043 $ & $0.224$ & $20.882\pm0.022 $ & $ 0.905 $ & $ 24.503\pm0.041 $\\

         0.016 & $ 14.751\pm0.059 $ & $ 0.065$ & $ 17.965\pm0.049 $ & $0.261$ & $21.231\pm0.020 $ & $ 1.058 $ & $ 24.893\pm0.073 $\\

         0.019 & $ 15.209\pm0.077 $ & $ 0.075 $& $ 18.294\pm0.050 $ & $0.305$ & $ 21.663\pm0.021 $ & $ 1.235 $ & $ 25.433\pm0.117 $\\

         0.022 & $ 15.482\pm0.053 $ & $ 0.088 $& $ 18.695\pm0.041 $ & $0.356$ & $22.042\pm0.022 $ & $ 1.443 $ & $ 25.572\pm0.152 $\\

         0.025  & $ 15.839\pm0.041 $ & $ 0.103 $ & $  19.094\pm0.032 $ & $ 0.416 $ & $ 22.469\pm0.028 $ & $ 1.686 $ & $ 26.248\pm0.222 $\\

         0.030  & $ 16.204\pm0.041 $ & $ 0.120 $ & $ 19.386\pm0.026 $ & $ 0.486 $ & $ 22.836\pm0.030 $ & $ 1.969 $ & $ 26.125\pm0.295 $\\

         0.035  & $ 16.550\pm0.036 $ & $ 0.140 $ & $ 19.825\pm0.025 $ & $ 0.568 $ & $ 23.272\pm0.031 $ & $ 2.300 $ & $ 26.967\pm0.294 $\\

    \hline
    \end{tabular}
\end{center}
\end{table}

\begin{table}
  \caption{the values of $\overline{m}_{\rm B}$ at the redshifts of BAO measurements obtained from two methods.}\label{Tab3}
   \begin{center}
\addtolength{\leftskip} {-1.5cm}
    \begin{tabular}{cccccc}
    \hline
    \hline
       ${\rm method}$  & $0.38$ & $0.51$ & $0.70$ & $0.85$ & $1.48$  \\
    \hline
     ${\rm I }$ & $22.218\pm0.024$ &$22.969\pm0.030$ & $22.787\pm0.040$ & $24.331\pm0.039$ & $25.682\pm0.164$ \\
    \hline
     ${\rm II}$ & $22.236\pm0.045$ &$22.990\pm0.044$ & $23.824\pm0.061$ & $24.317\pm0.071$ & $25.785\pm0.180$ \\
    \hline
    \end{tabular}
\end{center}
\end{table}

 In order to see whether the conclusions are changed by a different method and improve the robustness of the conclusion, we then use Artificial Neural Network (ANN) to reconstruct the $m_{\rm B}(z)$ function from the observation data and obtain the value of $m_{\rm B}$ from the reconstructed function to overcome the redshift-matching problem at high redshifts.  In general terms, ANN is a Deep Learning algorithm consisting of 3 layers - Input, Hidden and Output. The input layer has $n$ nodes, each of which inputs an independent variable, followed by the $m$ linked hidden layers and the output layer with activation function nodes in the basic architecture ~\citep{Schmidhuber15}.  By using adam optimization~\citep{Kingma14}, ANN estimates the error gradient from observations in the training dataset, and then updates the model weights and bias estimates during back propagation process to iterate toward an optimal solution. The process can be conveniently described in a vectorized way~\citep{Wang20a},

\begin{equation}
\bm{z}_{i+1}=\bm{b}_{i+1}+\bm{x}_i\bm{W}_{i+1}\,,
\end{equation}
\begin{equation}
\bm{x}_{i+1}=f(\bm{z}_{i+1})\,,
\end{equation}
where $\bm{x}_i$ is the input vector at the $i$th layer,  $\bm{b}_{i+1}$, and $\bm{W}_{i+1}$  are the offset vector and linear weights matrix with learnable parameter elements, $\bm{z}_{i+1}$ is the output vector after linear transformation of $\bm{x}_i$, and $f$ is the Exponential Linear Unit (ELU) activation function~\citep{Clevert15}, which has the form

\begin{align}\label{eq:elu}
&f(x) =
\begin{cases}
x & x > 0 \\
\alpha (\exp(x)-1) & x \leq 0
\end{cases} \, ,
\end{align}
where $\alpha$ denotes the hyper-parameter that controls the value to which an ELU saturates for negative net inputs and is set to 1~\citep{Wang20a}.

The key of  ANN is to find the function $f_{W,\mathbf{b}}$ which makes its output $f_{W,\mathbf{b}}(\mathbf{x})$ as close to the target value  $\mathbf{y}$ as possible. To achieve this goal, the difference between the predicted value $f_{W,\mathbf{b}}(\mathbf{x})$ of the current network and the target value $\mathbf{y}$, equal to the defined loss function  $\mathcal{L}$, should be minimum. During the training and evaluation strategy, the weight matrix of each layer is updated constantly to minimize  $\mathcal{L}$. Depending on the gradient descent method used, ANN reduces the loss value by continuously moving the loss value in the opposite direction of the current corresponding gradient. Formally, in a vectorized way~\citep{LeCun12}

\begin{eqnarray} \nonumber
\frac{\partial \mathcal L}{\partial z_{i + 1}}&=f^{\prime}(z_{i + 1})\frac{\partial \mathcal L}{\partial x_{i+1}}\,,\\ \nonumber
\frac{\partial \mathcal L}{\partial W_{i + 1}}&=x_{i}^T\frac{\partial \mathcal L}{\partial z_{i+1}}\,,\\ \nonumber
\frac{\partial \mathcal L}{\partial x_{i}}&= W_{i+1}^T \frac{\partial \mathcal L}{\partial z_{i + 1}}\,, \\
\frac{\partial \mathcal L}{\partial {b}_{i+1}} &= \left(\frac{\partial \mathcal L}{\partial z_{i+1}}\right)\,,
\end{eqnarray}
where the operator $\partial$ stands for partial derivatives, and $f^{\prime}$ is the derivative of  the nonlinear function $f$.

\begin{figure}
\centering
\includegraphics[width=0.5\textwidth, height=0.35\textwidth]{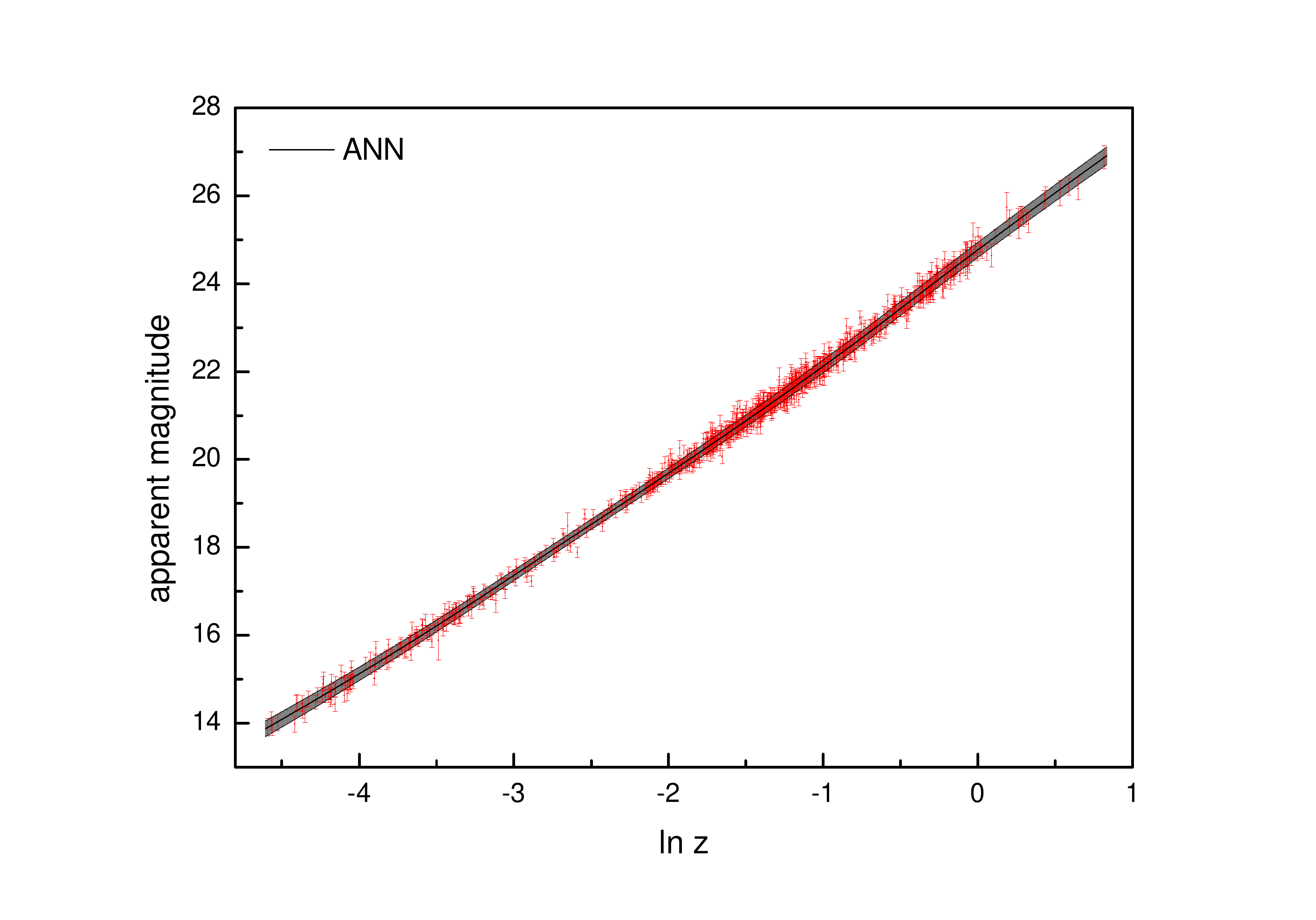}
\caption{ The reconstructed function $m_{\rm B}(z)$ with corresponding $1\sigma$ errors by using ANN~(black line), and the exact measurements of apparent magnitude in Pantheon sample~(red).}
\label{Fig2}
\end{figure}

Using the publicly released code called Reconstruct Functions with ANN (ReFANN) \footnote{https://github.com/Guo-Jian-Wang/refann}~\citep{Wang20a}, which explicitly describe the ANN method, we reconstruct the apparent magnitude-redshift relation as a function of logarithmic redshift~$\left( \rm lnz\right)$ and show the function with the estimation of $1\sigma$ confidence region in Figure~\ref{Fig2}. From Figure~\ref{Fig2}, it is easy to see that the uncertainties of the reconstructed function by ANN are almost equal to that of the observations, and the $1\sigma$ confidence region reconstructed by ANN can be considered as the average level of observational error.  We refer the reader to Ref.~\citep{Wang20a} for further details on this issue. Therefore,  it is difficult to achieve a robust CDDR test by using the $m_{\rm B}$ values  obtained from the reconstructed function directly. Given that the sample size of Pantheon with redshift $z<1$ is sufficient to apply the binning method~\citep{Meng12}, we bin the actual observational sample by taking $\lvert z_{\rm SNIa} - z_{\rm BAO}\rvert<0.005$ to obtain the values of $m_{\rm B}$. Due to the limited sample size of Pantheon at the redshift around $z=1.48$ and the failure of binning method, the value with $m_{\rm B}=25.785\pm0.180$ derived from the reconstructed function is used. This method is denoted as `II'. We summarize the values obtained by the method of combining the binned SNIa and ANN at the Table~\ref{Tab3}.

Using the measurements of $D_{\rm M}/r_{\rm d}$ given in Table~\ref{Tab1} and $\overline{m}_{\rm B}$ in Table~\ref{Tab3}, we can obtain the best-fitting CDDR parameters~$\eta_i$ and the confidence regions by minimizing the following $\chi^2$ function,

\begin{equation}\label{chi2}
\chi^2 =\sum^{5}_{i=1}\frac{{\Delta\mu_i}^2}{s^2_i}\,,
\end{equation}
where $\Delta\mu_i = \Delta_i- \mathcal{M} = 5\mathrm{log_{10}}[D_{\rm M}/r_{\rm d}(1+z)\eta(z)]- \overline{m}_{\rm B} - \mathcal{M}$ with $\mathcal{M}=M_{\rm B}-5\mathrm{log_{10}}r_{\rm d,f}-25$,  $s^2_i=\sigma^2_{\mu_{i,{\rm BAO}}}+\sigma^2_{\mu_{i,{\rm SN}}}$ with $\sigma_{\mu_{\rm BAO}}=5\sigma_{D_{\rm M}/r_{\rm d}}/\left(D_{\rm M}/r_{\rm d}\mathrm{ln}10\right)$, and $\sigma_{\mu_{\rm SN}}=\sigma_{\overline{m}_{\rm B}}$. In our analysis, we marginalize analytically the likelihood function over $\mathcal{M}$, the combination of $M_B$ and $r_{\rm d,f}$, with the method proposed in~\citep{Conley11} by assuming a flat prior on $\mathcal{M}$. And the marginalized $\chi^2$ is written as

\begin{equation}
\chi^2_{\rm marg}=a-\frac{b^2}{f}+\mathrm{ln}\frac{f}{2\pi}\,,
\end{equation}
where $a=\sum\limits_{i=1}^{5}\Delta^2_i/s^2_{i}$, $b=\sum\limits_{i=1}^{5}\Delta_i/s^2_{i}$ and $f=\sum\limits_{i=1}^{5}1/s^2_{i}$. Here, it should be pointed out that although the values of binned apparent magnitude are correlated, it is difficult to obtain correlations of the values of $\overline{m}_{\rm B}$ used in the tests. Given that the value of $\sigma_{\mu_{\rm BAO}}$ used in Equation~\ref{chi2} is greater than the one of $\sigma_{\overline{m}_{\rm B}}$ at the same redshift and the values of $D_{\rm M}/r_{\rm d}$ are not correlated except the two data points at low redshift, we have therefore ignored the correlation of all data points in the analysis.

\section{Results}\label{S3}

\begin{table}
  \caption{Summaries of the 68\% limits of $\eta_1$, $\eta_2$ and $\eta_3$ from the low-redshift data  points~(Sub) and the all data  points~(All) by using two methods.}\label{Tab4}
   \begin{center}
\addtolength{\leftskip} {-1.5cm}
     \begin{tabular}{cccc}
    \hline
    \hline
       $\mathrm{method\, I}$  & $\eta_1$ & $\eta_2$ & $\eta_3$   \\
    \hline
     Sub & $-0.052^{+0.085}_{-0.077} $ &$ -0.137^{+0.203}_{-0.177}$ & $ -0.085^{+0.132}_{-0.118}$ \\
    \hline
     All & $ -0.064^{+0.057}_{-0.052}$ &$ -0.181^{+0.160}_{-0.141}$ & $-0.110^{+0.097}_{-0.088} $ \\
    \hline
     $\mathrm{method\, II}$  & $\eta_1$ & $\eta_2$ & $\eta_3$   \\
    \hline
     Sub & $-0.037^{+0.110}_{-0.097} $ &$ -0.101^{+0.269}_{-0.225}$ & $ -0.061^{+0.173}_{-0.149}$ \\
    \hline
     All & $ -0.039^{+0.070}_{-0.062}$ &$ -0.119^{+0.207}_{-0.176}$ & $-0.070^{+0.122}_{-0.107} $ \\
    \hline
    \end{tabular}
\end{center}
\end{table}

Since that the low redshift~$(z<1)$ SNIa plus BAO data have been used to test the validity of CDDR in the literatures~\citep{Wu15,Lin18,Xu20}, for a comparison, we perform the constraints using the first four data points~(Sub) and all five data points~(All) listed in Table~\ref{Tab3}, respectively, in order to test the constraint ability of high redshift BAO measurement on the CDDR.

\begin{figure}
\centering
\includegraphics[width=0.315\textwidth]{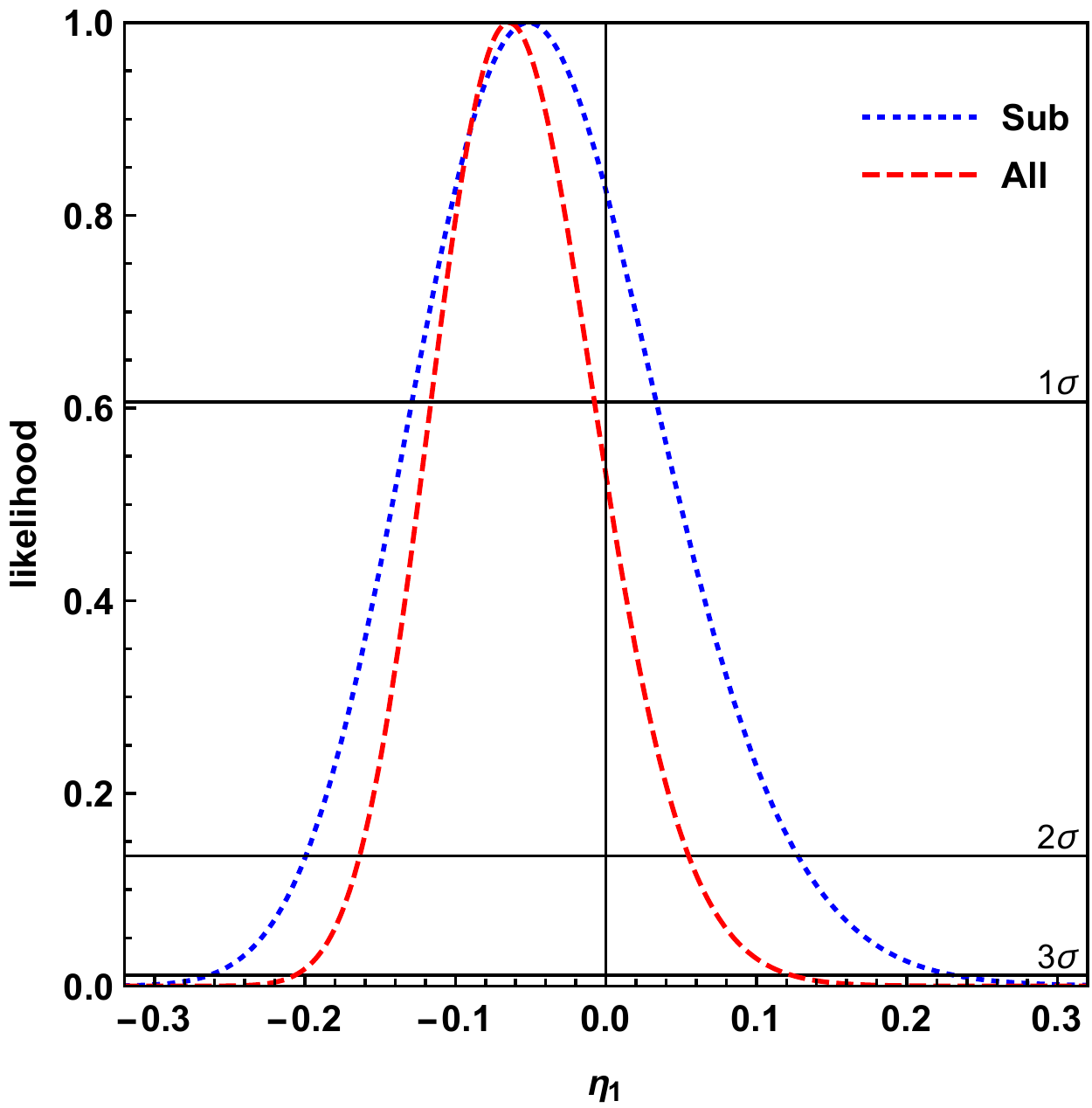}
\includegraphics[width=0.315\textwidth]{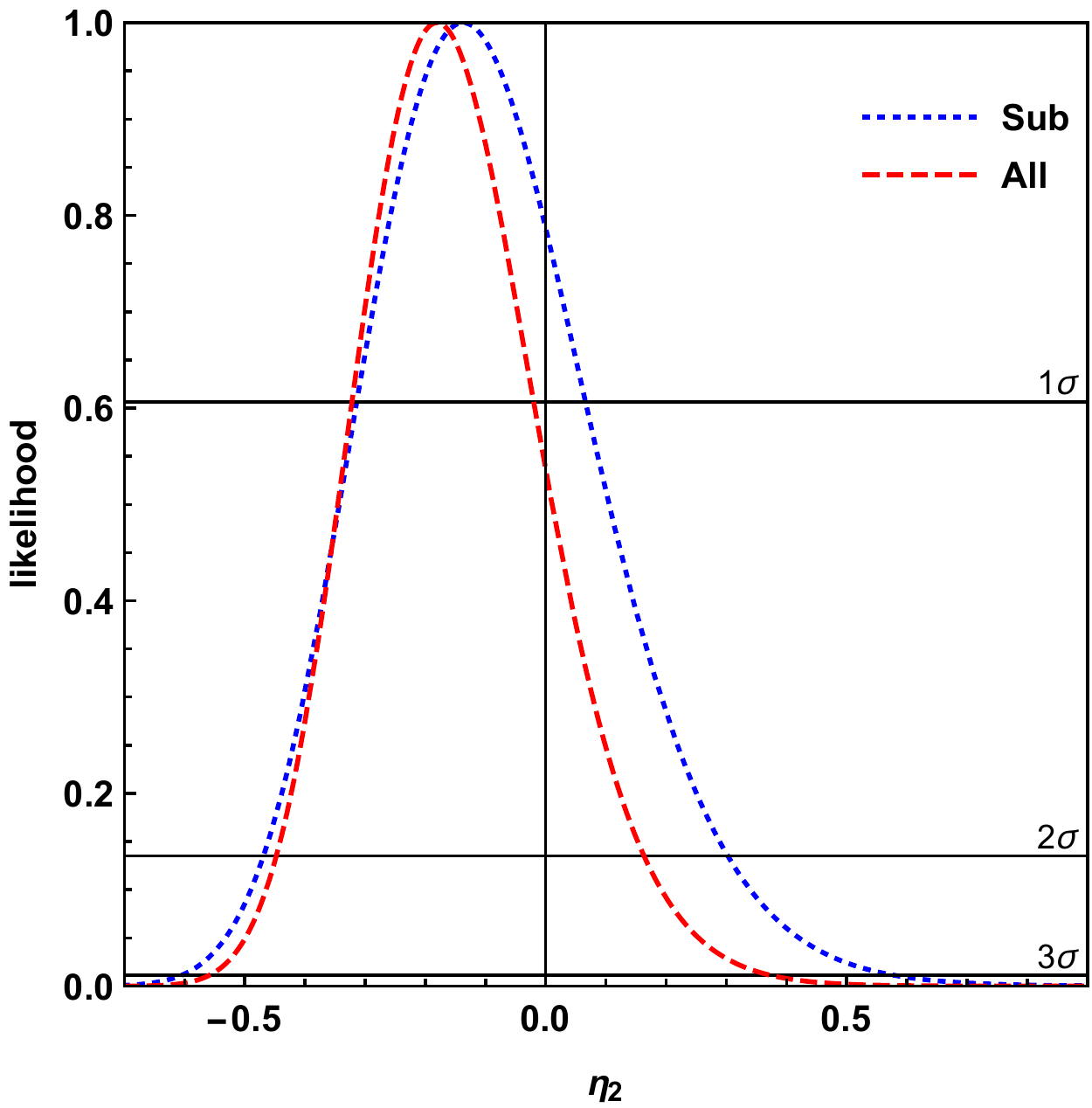}
\includegraphics[width=0.315\textwidth]{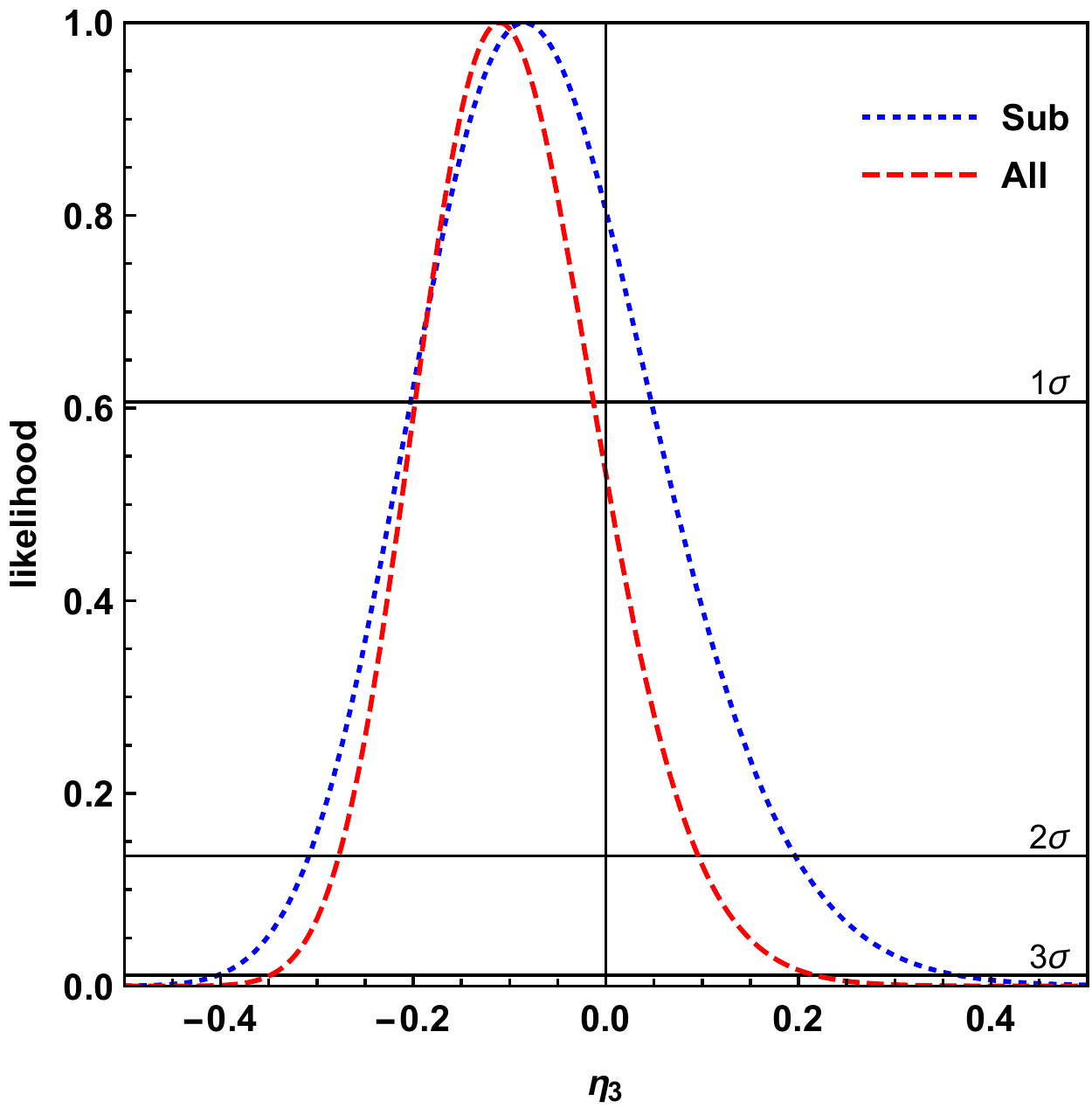}
\caption{The likelihood distributions of $\eta_1$, $\eta_2$ and $\eta_3$ from the low-redshift data points~(Sub) and the all data  points(All) by using method I.}
\label{Fig3}
\end{figure}

\begin{figure}
\centering
\includegraphics[width=0.315\textwidth]{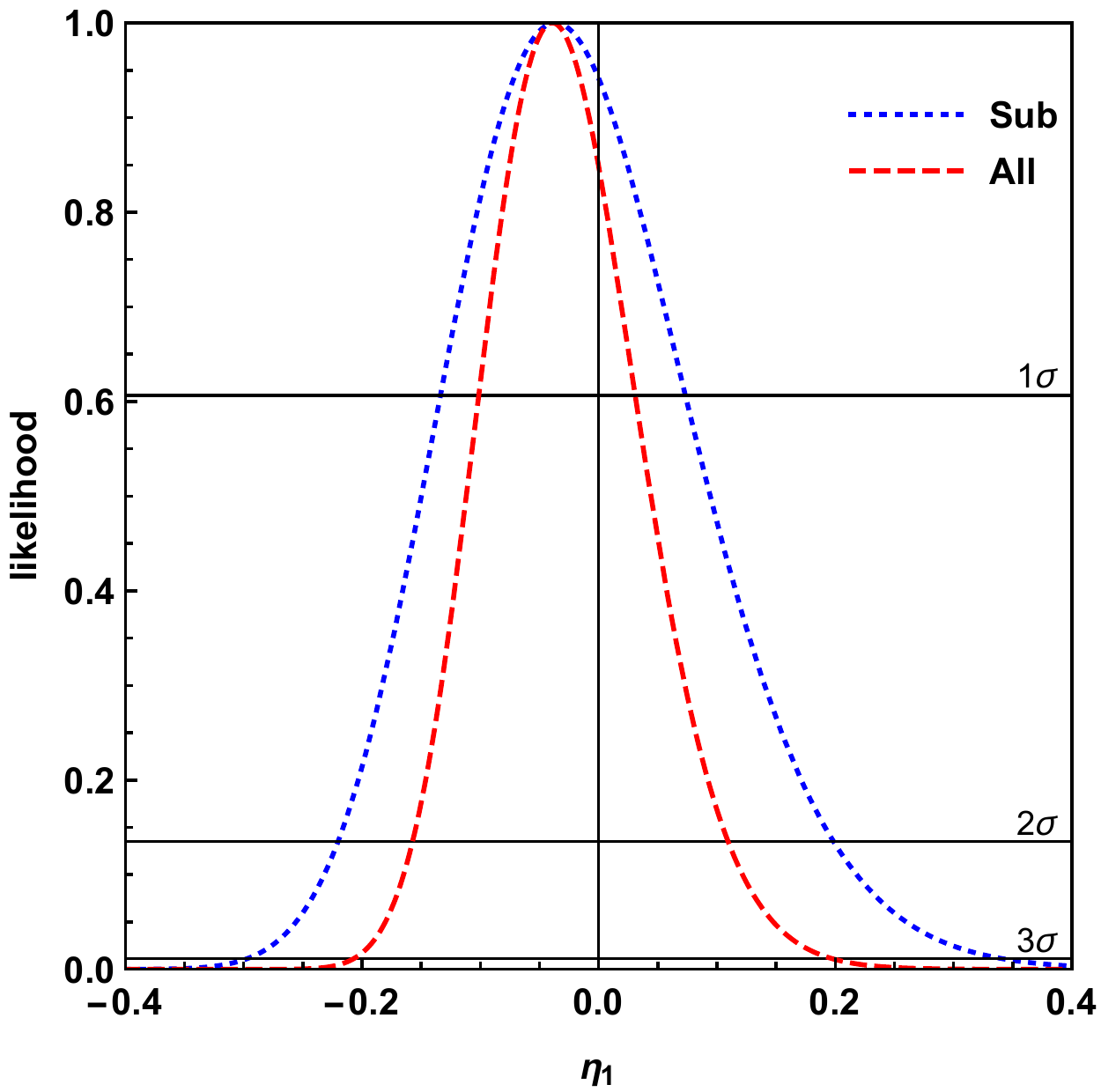}
\includegraphics[width=0.315\textwidth]{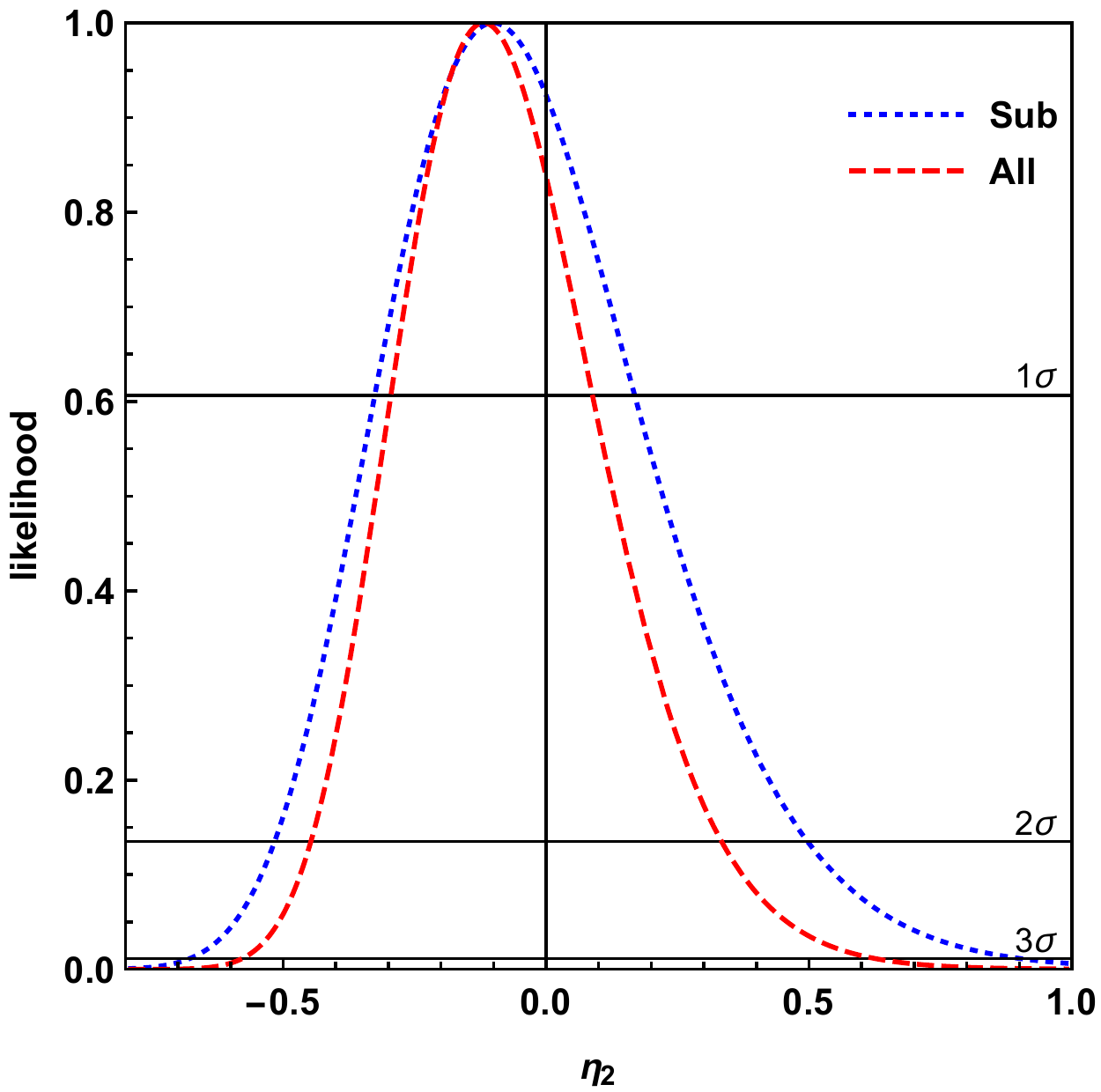}
\includegraphics[width=0.315\textwidth]{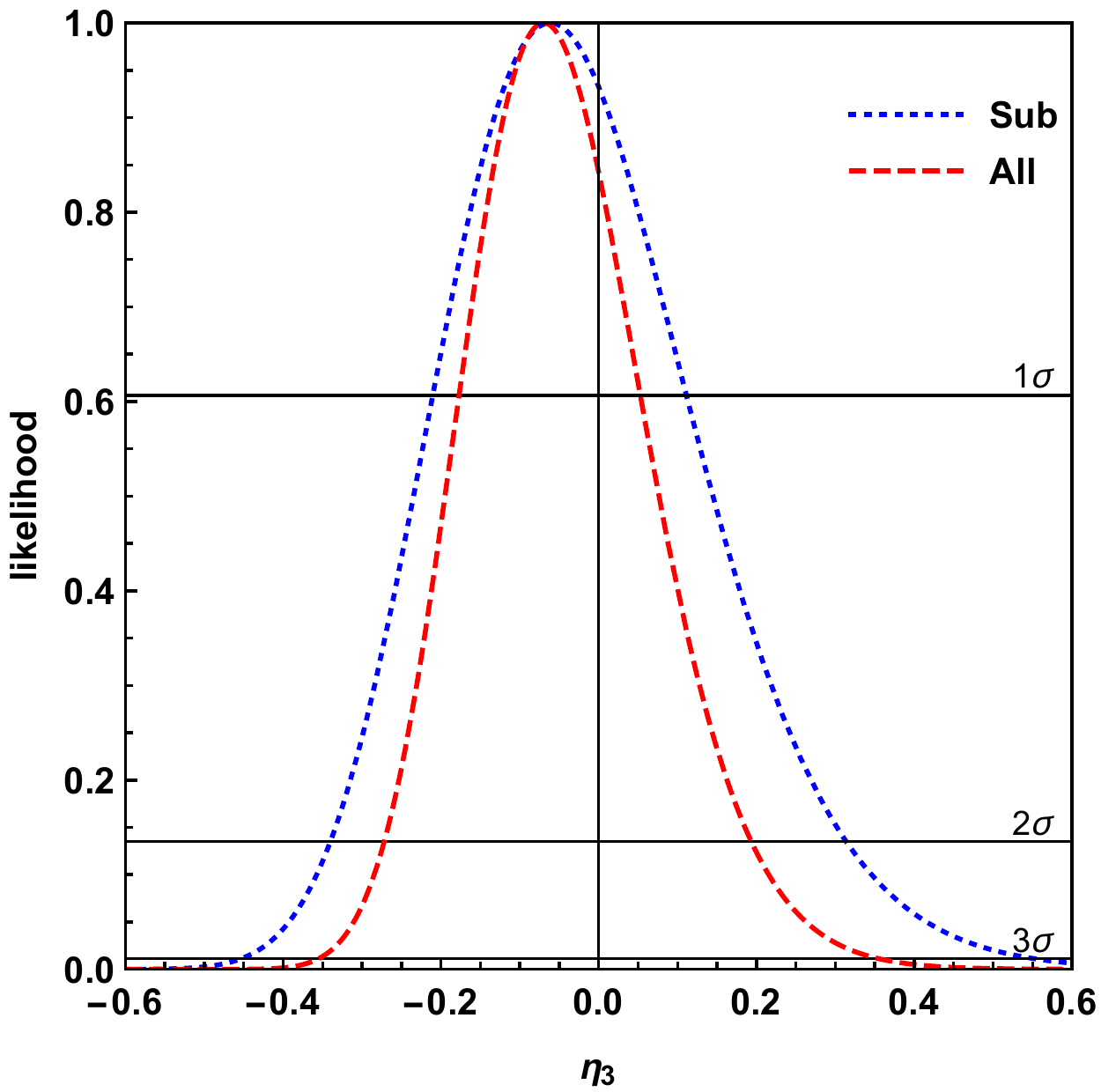}
\caption{The likelihood distributions of $\eta_1$, $\eta_2$ and $\eta_3$ from the low-redshift data points~(Sub) and the all data  points(All) by using method II.}
\label{Fig4}
\end{figure}

We first focus on the results obtained using method I and show the marginalized likelihood distributions of CDDR parameters $\eta_i$ are shown in Figure~\ref{Fig3}  with the corresponding 68\% limits summarizing in Table~\ref{Tab4}. From Figure~\ref{Fig3} and Table~\ref{Tab4}, one can see that when the low redshift data is used, the CDDR is valid at $1\sigma$ CL for all three parameterizations with $\eta_1=-0.052^{+0.085}_{-0.077}$, $\eta_2=-0.137^{+0.203}_{-0.177}$, $\eta_3=-0.085^{+0.132}_{-0.118}$. However, once the data point at redshift $z=1.48$ is combined, the different results are obtained. In this case, $\eta_1=-0.064^{+0.057}_{-0.052}$, $\eta_2=-0.181^{+0.160}_{-0.141}$, $\eta_3=-0.110^{+0.097}_{-0.088}$, respectively, which means that the CDDR is valid at $2\sigma$ CL for all three parameterizations. This suggests that the tests of CDDR are not sensitive to the parametrization of $\eta(z)$, and there is no obvious evidence for violation of the CDDR. In addition, it is easy to see that by comparing the results obtained from the low redshift data with $\Delta\eta_1=0.162$, $\Delta\eta_2=0.380$, $\Delta\eta_3=0.250$, there is a significant improvement on the constraints on $\eta_i$ with $\Delta\eta_1=0.109$, $\Delta\eta_2=0.301$, $\Delta\eta_3=0.185$, and the confidence intervals are reduced by 33\%, 21\%, 26\% for the three parameterizations respectively when the high redshift data is added.

We then turn to the results in Figure~\ref{Fig4} and Table~\ref{Tab4} obtained by using method II. Similar to the results obtained by using method I, the CDDR is consistent well with the SNIa and BAO observations. From the low redshift data, $\eta_1=-0.037^{+0.110}_{-0.097}$, $\eta_2=-0.101^{+0.269}_{-0.225}$, $\eta_3=-0.061^{+0.173}_{-0.149}$, and $\eta_1=-0.039^{+0.070}_{-0.062}$, $\eta_2=-0.119^{+0.207}_{-0.176}$, $\eta_3=-0.070^{+0.122}_{-0.107}$  from all data. In other words, the high redshift data improves the constraints on the violation parameters from $\Delta\eta_1=0.207$, $\Delta\eta_2=0.494$, $\Delta\eta_3=0.322$ to $\Delta\eta_1=0.132$, $\Delta\eta_2=0.383$, $\Delta\eta_3=0.229$ with the precisions increasing by 36\%, 22\% and 29\% corresponding to the first, second and third parameterized forms. In addition, we can see that each $1\sigma$ error is obviously larger than that from the method I. This is the advantage of method I that the observations can make a more rigorous constraint on the CDDR as more SNIa data are used in this method to derive more precise values of $m_{\rm B}$ at the BAO redshifts.

In order to highlight the constraint ability of the newest BAO measurements, it is necessary to compare our results with the previous constraints on $\eta_i$ from different data sets of SNIa and BAO. Combining five BAO data covering the redshift $0.44\leq z \leq 0.57$  with the binned SNIa data in the range $\Delta z =|z_{\rm BAO}-z_{\rm SNIa}| < 0.005$ from Union2.1 sample, \citet{Wu15} tested the CDDR with the first two parameterizations used in this paper. They found that when the dimensionless Hubble constant $h$ is marginalized with a flat prior, the constraints on $\eta_i$ is very weak with $\eta_1=-0.174^{+0.253}_{-0.199}$ and $\eta_2=-0.409^{+0.529}_{-0.381}$. Using the latest Pantheon sample and BOSS DR12 BAO measurements covering the redshift $0.31\leq z \leq 0.72$, \citet{Xu20} updated the constraints, and found that $\eta_1=-0.07\pm0.12$, $\eta_2=-0.20\pm0.27$ and $\eta_3=-0.12\pm0.18$, which revealed more stringent constraints. Furthermore, \citet{Lin18} tested the CDDR by combining the ADD data from galaxy clusters and BAO measurements with the Pantheon sample. In their work, besides the five BAO data used in~\citep{Wu15}, the authors added one BAO data point at redshift $z=2.34$. However, they found that the results are almost unaffected when the high redshift BAO data is added, because that the reconstructed $\mu - z$ function with the Gaussian processes has large uncertainty at high redshift, and the best-fitting values are $\eta_1=-0.04\pm0.12$ and $\eta_2=-0.05\pm0.22$. Through these comparisons, one can conclude that newest BAO data, especially the eBOSS DR16 data at effective redshift $z=1.48$, could effectively strengthen the constraints on the violation parameters of CDDR.

Finally, it is worth comparing our results with some tests that have been already studied by using other high-redshift astrophysical probes to obtain ADD previously. In particular, many efforts have been made to perform robust tests of CDDR by combining the ADDs derived from ultra-compact structure in radio quasars~\citep{Cao17} with  the LDs obtained from Pantheon SNIa sample~\citep{He22}, the relation between the UV and X-ray luminosities of quasars~\citep{Zheng20}, observations of HII galaxies~\citep{Liu21}, and simulated gravitational wave data~\citep{Qi19}, respectively. In these works, the authors tested CDDR in the redshift range $z>2$, and found that the validity of CDDR was in good agreement with the current observational data. More notably, their results showed that these combinations of observational data could derive robust constraints on the violation parameter at the precision of $10^{-2}$ or higher, whereas in our work, the constraint results obtained from current SNIa and BAO data do not achieve such accuracy. However, given that those works all use simulated data to solve the redshift matching problem, while we directly use five data points derived from the actual observational data, we can expect a better validity check of the CDDR as more high-precision, high-redshift observations of available BAO and SNIa.


\section{Conclusion and Discussions}\label{Sec:Sec4}
The CDDR plays a fundamental role in astronomical observations and modern cosmology, while it may be violated if one of the assumptions underlying this relation is not true. In this paper, we have proposed a new model-independent test for the CDDR with the Pantheon SNIa sample and the newest BAO measurements including the eBOSS DR16 quasar sample at effective redshift $z=1.48$.  In our analysis, three parameterized forms $\eta\left(z,\eta_1\right)=1+\eta_1z$, $\eta\left(z,\eta_2\right)=1+\eta_2z/(1+z)$, and $\eta\left(z,\eta_3\right)=1+\eta_3{\rm ln}(1+z)$ are used to describe the possible violation of CDDR.  In particular, two methods are used to derive the values of $m_{\rm B}$ at the redshifts of BAO measurements to overcome the redshift-matching problem. We first provide a compressed form of the Pantheon SNIa sample by using a piecewise linear function of $\mathrm{ln}(z)$, which shows that it still remains accurate for cosmological constraints, to derive the values of $m_{\rm B}$ at the redshifts of BAO measurements. We then obtain the values by using the binning SNIa method at redshift $z<1$ and the ANN method at $z=1.48$.

The results show that the tests of CDDR are not sensitive to the parametrization of $\eta(z)$. And there is no obvious violation of the CDDR. Moreover, it is observed that the high redshift BAO and SNIa data can effectively strengthen the constraints on the parameters of CDDR. The confidence intervals of the violation parameter in the three  parameterizations are decreased by 33\%, 21\%, 26\% in the framework of first method, while they are reduced by 36\%, 22\%, 29\% in the second method, respectively. Furthermore, by comparing the results obtained from the two methods, we find that applying the method of compressed form of Pantheon sample, we can use more actual SNIa data to derive a more precise value of $m_{\rm B}$ and thus get more rigorous constraints on the violation parameter.

As the final remarks, due to the lack of high redshift SNIa sample, the test of CDDR with BAO and SNIa data is hard to reach a higher redshift, though eBOSS collaboration has release one other high redshift data, i.e. the measurement with Ly$\rm\alpha$ absorption and quasars from BOSS and eBOSS at an effective redshift $z=2.33$~\citep{Bourboux20}. We can expect that the validity of the CDDR will be better checked with the more accurate and wider redshift range measurements of BAO and SNIa from the further observations, such as the Euclid Satellite~\citep{Euclid} and Dark Energy Spectroscopic Instrument~(DESI,~\citet{DESI}). Furthermore, the method of compressed form of observations proposed in this paper presents a new idea of overcoming the redshift-matching problem in high redshift range. We therefore expect that the method can be extended to other high redshift standard candles with limited sample size, such as gamma ray burst and HII galaxies, to test the CDDR, which is an interesting topic for future investigation.

\acknowledgments

We thank the anonymous reviewer for his/her very enlightening comments. We are very grateful to Guojian Wang  for his introduction of the ReFANN code. This work was supported by the National Natural Science Foundation of China under grants Nos. 11505004, 11865018, and 12192221, and the Anhui Provincial Natural Science Foundation of China (1508085QA17).

\bibliography{}

\begin{sidewaystable}
\appendix
\section{The covariance matrix of binned apparent magnitude parameters}
The covariance matrix of 36 binned apparent magnitude  parameters $m_{\rm B,b}$ obtained by running CosmoMC is given in Table~\ref{TabA}.
\setcounter{table}{0}
\renewcommand{\thetable}{A\arabic{table}}
  \caption{Covariance matrix of the binned apparent magnitudes.}\label{TabA}
  \begin{center}
  \resizebox{1\columnwidth}{!}{
\addtocounter{MaxMatrixCols}{36}
 $ \begin{pmatrix}
 $203884$ & $-52108$ & $20037$ & $-370$ & $	4708 $ & $2351 $ & $2861 $ & $	2619 $	& $2786 $ & $	3303 $ & $	2692$ & $ 	2675 $ & $	561 $ & $	2694 $ & $	-909 $ & $-405 $ & $	-282 $ & $-1082 $ & $-777 $ & $	-943 $ & $	-414 $ & $	-144 $ & $	-32 $ & $	-841 $ & $	-1417 $ & $	291 $ & $	-1732 $ & $-911 $ & $	-1627 $ & $	-1142 $ & $	680 $ & $	572 $ & $3094 $ & $	3949 $ & $	2467 $ & $	4738$ \\
$ $ & $173352 $ & $	-43188 $ & $17031 $ & $	-5196 $ & $	5404$ & $2332$ & $ 2652$ & $ 	2488 $ & $3113 $ & $1905 $ & $2768$ & $	1324 $ & $797$ & $ -962 $ & $-482 $ & $-186 $ & $-650 $ & $-1513 $ & $-964$ & $ -525$ & $ 452 $ & $-394$ & $ -382 $ & $-1286 $ & $56 $ & $-2119$ & $ -1080$ & $ -1784$ & $ -487$ & $ 879$ & $ 1421$ & $ 1311$ & $ 2460 $ & $4089$ & $ -276$\\
$ $ & $ $ & $77426$ & $ -17286$ & $ 14544$ & $ -167$ & $ 3000$ & $ 2258$ & $ 2767 $ & $	3767 $ & $1849$ & $ 2504 $ & $1318 $ & $	2469$ & $ -894 $ & $-270 $ & $-139$ & $ -803$ & $ -1051 $ & $	-871 $ & $-427$ & $ -275 $ & $52 $ & $-752 $ & $	-1044 $ & $	54 $ & $	-1961$ & $ -531 $ & $-2295$ & $ -758 $ & $	-316 $ & $395$ & $ 675 $ & $4616 $ & $-452 $ & $	4355 $\\
$ $ & $ $ & $ $ & $ 34537$ & $ -14612 $ & $	8038 $ & $1414$ & $ 2761 $ & $2200 $ & $	2337 $ & $1921 $ & $2215 $ & $1179 $ & $	1554 $ & $-632 $ & $-572 $ & $-464 $ & $	-760 $ & $-1322 $ & $-976 $ & $	-649 $ & $	13 $ & $-435 $ & $-393 $ & $-1001 $ & $	33 $ & $	-1623 $ & $	-1351 $ & $	-1600 $ & $	-354 $ & $872 $ & $594 $ & $	2828 $ & $2235 $ & $	4016$ & $ 2628$ \\
$ $ & $ $ & $ $ &$ $ & $58692 $ & $	-6054 $ & $4238 $ & $	1148 $ & $2185 $ & $1128 $ & $	772 $ & $939 $ & $115 $ & $	862 $ & $-1425$ & $ -1031 $ & $-423 $ & $	-702 $ & $-1129 $ & $	-1026 $ & $	-731 $ & $-144 $ & $-421 $ & $	-196 $ & $-857 $ & $176 $ & $-712 $ & $ -1201$ & $ -846 $ & $-401 $ & $	1169 $ & $	1583 $ & $1373 $ & $3621 $ & $1674 $ & $	2137$ \\
$ $ & $ $ & $ $ & $ $ & $ $ & $28229 $ & $	-3112$ & $ 3304 $ & $1613 $ & $	1225 $ & $	1142 $ & $1169 $ & $443 $ & $979 $ & $	-1034$ & $ -728 $ & $-414 $ & $	-688 $ & $	-1138 $ & $	-930 $ & $-650$ & $ -129 $ & $	-448 $ & $-275$ & $ -691 $ & $87 $ & $-1228 $ & $	-950 $ & $-875$ & $ -226 $ & $1209$ & $ 	1011 $ & $2292 $ & $2750 $ & $	3186 $ & $	3863$ \\
$ $ & $ $ & $ $ & $ $ & $ $ & $ $ & $16952$ & $ -2567 $ & $3003 $ & $	1863 $ & $	1288 $ & $1681 $ & $	846 $ & $1168$ & $ -614 $ & $-214$ & $ -240 $ & $ -551 $ & $-773 $ & $	-449$ & $ -458 $ & $-52$ & $ -128 $ & $	-601$ & $ -760 $ & $31 $ & $-1226 $ & $-411 $ & $	-1399 $ & $	-502 $ & $889 $ & $	-20 $ & $	2123 $ & $2599 $ & $2211 $ & $3547$ \\
$ $ & $ $ & $  $ & $ $ & $ $ & $ $ & $ $ & $ 	16613$ & $-2115 $ & $2756 $ & $	632 $ & $	1008 $ & $511 $ & $	870 $ & $-450 $ & $	-253$ & $ -365 $ & $-507 $ & $-916 $ & $-615$ & $ 	-383 $ & $	52 $ & $-31 $ & $-356 $ & $	-810$ & $ 54 $ & $-1231 $ & $	-346 $ & $	-1326 $ & $-206 $ & $	698 $ & $1210 $ & $	2498 $ & $	2907 $ & $	3060 $ & $	3295$ \\
$ $ & $ $ & $ $ & $ $ & $ $ & $ $ & $ $ & $ $ & $ 13173 $ & $	-1131 $ & $	1980 $ & $1201 $ & $	522 $ & $1154 $ & $	-696 $ & $-199 $ & $	-327 $ & $-666 $ & $-910 $ & $-690 $ & $	-497 $ & $-36 $ & $	-123 $ & $-469 $ & $	-729 $ & $168 $ & $	-889 $ & $-424 $ & $	-1127 $ & $	-239 $ & $931 $ & $	663 $ & $	1443 $ & $2335 $ & $2143 $ & $1982$ \\
$ $ & $ $ & $ $ & $ $ & $ $& $ $ &	$ $ & $ $ & $ $ & $21518$ & $ -1913$ & $ 4007$ & $ 	1419$ & $ 2100 $ & $-481 $ & $-316 $ & $	-261 $ & $-545$ & $ -950 $ & $-658 $ & $	-458 $ & $-106$ & $ -352 $ & $-414 $ & $-852$ & $ -83 $ & $-988 $ & $	-547 $ & $-1415$ & $ 	-583$ & $ 214$ & $ 485$ & $ 679 $ & $2047$ & $ 1649 $ & $1614 $\\
$ $ & $ $ & $ $ &$ $ & $ $ & $ $ & $ $ & $ $ & $ $ & $ $ & $24705 $ & $-492 $ & $1642 $ & $1465 $ & $	-834 $ & $-713$ & $ 112 $ & $-472$ & $ -595 $ & $-418 $ & $	-424 $ & $	-88 $ & $-420 $ & $	153 $ & $-901 $ & $	222$ & $ -483 $ & $-1601 $ & $-425 $ & $	-487 $ & $-288$ & $ -1973 $ & $-307$ & $ -2116 $ & $	-2838 $ & $	-2060 $\\
$ $ & $ $ & $ $ & $ $ & $ $ & $ $ & $ $ & $ $ & $ $ & $ $ & $ $ & $18883$ & $ -1433 $ & $	3199 $ & $-432 $ & $-314 $ & $186 $ & $	-341$ & $ -368 $ & $-755 $ & $-237 $ & $-130$ & $ 	-137 $ & $-433$ & $ -712 $ & $-163$ & $ 	-896 $ & $-730 $ & $-1297 $ & $	-1246 $ & $	-485 $ & $-265 $ & $293 $ & $-159 $ & $	-410 $ & $460 $\\
$ $ & $ $ & $ $ & $ $ & $ $ & $ $ & $ $ & $ $ & $ $ & $ $ & $ $ & $ $	& $23573$ & $ -3649 $ & $	196 $ & $-242 $ & $	60 $ & $-131$ & $ 	-332 $ & $-95 $ & $	-290 $ & $98 $ & $90 $ & $-475 $ & $	-276 $ & $-273 $ & $-523 $ & $	-302 $ & $-1185 $ & $-1050 $ & $-257 $ & $	190 $ & $1703 $ & $	-246$ & $ 386 $ & $	-234$ \\
$ $ & $ $ & $ $ & $ $ & $ $ & $ $ & $ $ & $  $ & $ $ & $ $ & $ $ & $ $ & $ $	& $ 25154 $ & $-3435$ & $577 $ & $	217 $ & $-608 $ & $	-368$ & $ -387 $ & $-102 $ & $277 $ & $117 $ & $	-455$ & $ -1144 $ & $-279 $ & $	-654 $ & $	126 $ & $-1863$ & $ -1126 $ & $	-631 $ & $	-1347 $ & $	-501 $ & $1101$ & $ 1353 $ & $	1584 $\\
$ $ & $ $ & $ $ & $ $ & $ $ & $ $ & $ $ & $ $ & $ $ & $ $ & $ $ & $ $ & $ $ & $ $ & $16493$ &  $ -1266 $ & $856$ & $ 343 $ & $903 $ & $	397 $ & $495 $ & $76 $ & $209 $ & $	164 $ & $	-109 $ & $-416 $ & $206 $ & $579 $ & $-235 $ & $	-855 $ & $-1491 $ & $-869 $ & $	-2292 $ & $-2046 $ & $-1905 $ & $	-3326 $\\
$ $ & $ $ & $ $ & $ $ & $ $ & $ $ & $ $ & $ $ & $ $ & $ $ & $ $ & $ $ & $ $ & $ $ & $ $ & $	10143 $ & $-970 $ & $	800 $ & $357 $ & $422$ & $ 455 $ & $	180 $ & $322 $ & $-123 $ & $	-61 $ & $-440 $ & $	-182 $ & $2 $ & $-717 $ & $-856 $ & $	-1140 $ & $	-1315 $ & $	-862 $ & $53 $ & $36 $ & $-527$ \\
$ $ & $ $ & $ $ & $ $ & $ $ & $ $ &	$ $ & $ $ & $ $ & $ $ & $ $ & $ $ & $ $ & $ $ & $ $ & $ $ & $6687 $ & $	-1148 $ & $	1021 $ & $111 $ & $301 $ & $188$ & $177$ & $ 114 $ & $-51 $ & $	-356 $ & $-49 $ & $	-123 $ & $-134 $ & $	-798 $ & $-1233 $ & $-807 $ & $	-1283 $ & $	-1327 $ & $	-1756 $ & $	-360 $\\
$ $ & $ $ & $ $ & $ $ & $ $ & $ $ &	$ $ & $ $ & $ $ & $ $ & $ $ & $ $ &	$ $ & $ $ & $ $ & $ $ & $ $ & $6295 $ & $	-888 $ & $1040$ & $ 	414 $ & $203 $ & $343 $ & $	185$ & $ 373 $ & $-512 $ & $	36 $ & $-224 $ & $	-202$ & $ 	-912 $ & $	-1885$ & $	-1833 $ & $	-2394 $ & $-2306$ & $ -2521 $ & $	-3224$ \\
$ $ & $ $ & $ $ & $ $ & $ $ & $ $ &	$ $ & $ $ & $ $ & $ $ & $ $ & $ $ & $ $ & $ $ & $ $ & $ $ & $ $ & $ $ & $6624 $ & $	-999 $ & $	1038 $ & $56 $ & $243 $ & $265 $ & $	510 $ & $	-554 $ & $174 $ & $	-246 $ & $	-9 $ & $	-1012 $ & $	-1852 $ & $	-987 $ & $-2536 $ & $	-1783 $ & $	-2572 $ & $	-2392 $\\
$ $ & $ $ & $ $ & $ $ & $ $ & $ $ &	$ $ & $ $ & $ $ & $ $ & $ $ & $ $ & $ $ & $ $ & $ $ & $ $ & $ $ & $ $ &	$ $ & $4840 $ & $-698 $ & $	504 $ & $205 $ & $287 $ & $	187 $ & $-372 $ & $56 $ & $-123 $ & $	-123 $ & $-626 $ & $	-1388 $ & $	-2277 $ & $	-1713 $ & $	-2848 $ & $-1476 $ & $-2627 $\\
$ $ & $ $ & $ $ & $ $ & $ $ & $ $ &	$ $ & $ $ & $ $ & $ $ & $ $ & $ $ &	$ $ & $ $ & $ $ & $ $ & $ $ & $ $ &	$ $ & $ $ & $5006 $ & $	-786 $ & $649 $ & $	98 $ & $250 $ & $-539 $ & $-248 $ & $	-484 $ & $-513 $ & $-822 $ & $	-1930 $ & $	-2462 $ & $	-3491 $ & $	-1662 $ & $-3183 $ & $-2022 $\\
$ $ & $ $ & $ $ & $ $ & $ $ & $ $ &	$ $ & $ $ & $  $	& $ $ & $ $ & $ $ &	$ $ & $ $ & $ $ & $ $ & $ $ & $ $ &	$ $ & $ $ & $ $ & $3932 $ & $	-607 $ & $	265 $ & $-25 $ & $-255 $ & $	-328 $ & $-195 $ & $-547 $ & $-799 $ & $	-1078 $ & $	-2075 $ & $	-2016 $ & $	-1970 $ & $-2032 $ & $-2180$ \\
$ $ & $ $ & $ $ & $ $ &	$ $ & $ $ &	$ $ &	$ $ &	$ $ & $ $ &	$ $ & $ $ & $ $ & $ $ &	$ $ &	$ $ & $ $ &	$ $ & $ $ &	$ $ & $ $ &	$ $ &	$4419$ & $-1124$ &	$509$ &	$-454$ 	&$-295$ &	$-213$ & $-622$  &$-621$ &	$-1236$ &	$-1467$ & $-1576$ &	$-1795$ & $-1104$ &	$-1291$ \\
$ $ & $ $ &	$ $ & $ $ &	$ $ & $ $ &	$ $ & $ $ & $ $ & $ $ & $ $ & $ $ & $ $ & $ $ & $ $ & $ $ &	$ $&  $ $ & $ $ & $ $ &	$ $ & $ $ &	$ $ &	$4684$ & $-941$ & $511$ & $325$&  $-238$ 	&$322$ & $-366$ & $-897$ & $-1832$ &	$-2795$ & $-2824$ &	$-3094$ & $-3059$ \\
$ $ & $ $ &	$ $ & $ $ &	$ $ & $ $ &	$ $ & $ $ &	$ $ & $ $ &	$ $  &$ $ &	$ $ & $ $ &	$ $ &	$ $ & $ $ &	$ $ & $ $ &	$ $ & $ $ &	$ $  & $ $ & $ $ & $7959$ &	$-1992$ & $1090$ &	$-366$ & $614$ & $-467$ & $-1329$ &	$-1507$ & $-3123$ & $-3675$ & $-2914$  &$-2453$ \\
$ $ &	$ $ &	$ $ &	$ $ &	$ $ &	$ $ &	$ $ 	&$ $ &	$ $ 	&$ $ &	$ $ &	$ $ 	&$ $ &	$ $ &	$ $ &	$ $ &	$ $ &	$ $ &	$ $ &	$ $ &	$ $ &	$ $ &	$ $ &	$ $ &	$ $ 	&$8938$ &	$-1264$&	$1473$ &	$300$ &	$916$ &	$1325$& 	$1124$ &	$1483$ &	$-422$ &	$668$ &	$581$ \\
$ $ &	$ $ &	$ $ &	$  $&	$ $ &	$ $ &	$ $ &	$ $ &	$ $ &	$ $ &	$ $ 	&$ $ &	$ $ &	$ $ &	$ $ &	$ $ &	$ $& 	$ $ 	&$  $	&$ $ &	$ $ &	$ $ &	$  $	&$   $&	$ $ &	$ $ &	$9366$&	$-2470$ &	$2386 $	&$337 $	&$327$ &	$123 $	&$-818 $&	$-1642$ &	$-1103$ &	$-1996$ \\
$  $	&$ $ &	$ $ &	$ $ &	$ $ &	$ $ &	$ $ &	$ $ &	$ $ &	$ $ &	$ $ &	$ $ &	$ $ 	&$ $ &	$ $ &	$ $ &	$ $& 	$ $ &	$ $ &	$ $ &	$ $ &	$ $& 	$ $ &	$ $ &	$ $ &	$ $ &	$ $ &	$17279$ &	$-3461$ &	$2239$& 	$489$ &	$2921$ &	$2683$ &	$2022$ &	$3483$ &	$2904$ \\
$ $ 	&$ $ &	$ $ &	$ $ 	&$ $ &	$ $ &	$ $ &	$ $ &	$ $ 	&$ $ &	$ $ &	$ $	&$ $ 	&$ $ &	$ $ &	$ $ &	$ $ &	$  $	&$ $ &	$ $ &	$ $ &	$ $ &$ $ &	$ $ &	$ $ &	$ $ &	$ $ &	$ $& 	$13398$ 	&$-1070$ &	$4028$ &	$1684$ &	$920$ &	$-251 $	&$926$ &	$852$ \\
$ $ &	$ $ &	$ $ &	$ $ &	$ $ &	$ $ &	$ $ 	&$ $ &	$ $ &	$ $ &	$  $&	$ $ &	$ $ 	&$ $ &	$ $ &	$ $ &	$ $& 	$ $ &	$ $ &	$ $ &	$ $ &	$ $ &	$ $ &	$  $	&$ $ &	$ $ &	$ $ &	$ $ &	$ $ &	$16903$ &	$-5699$ &	$7044$ &	$4796$ &	$6705$ &	$5288$ &	$5750$ \\
$  $ &	$ $ &	$ $&	$ $ &	$ $ &	$ $ &	$ $ &	$ $ &	$ $ &	$ $ &	$  $	&$ $ &	$ $ &	$ $ &	$ $ &	$ $ &	$ $ &	$ $ &	$ $ 	&$ $ &	$ $ &	$ $ 	&$ $ 	&$ $ &	$ $ &	$ $ &	$ $ &	$ $ &	$ $ &	$ $ &	$53196$ &	$4874$ &	$18735$ &	$10287$ &	$17295$ &	$14601$ \\
$ $ &	$ $&	$ $ &	$ $ &	$ $ &	$ $ &	$ $ 	&$ $ &	$ $ &	$ $&	$ $ &	$ $ 	&$ $ 	&$ $ &	$ $ &	$ $ &	$  $&	$  $	&$ $ 	&$ $ &	$ $ &	$  $&	$ $ 	&$ $ &	$ $ &	$ $ 	&$ $ &	$ $ &	$ $ &	$ $ &	$ $ &	$136322$ &	$-55347$ &	$66087$ &	$-3182$ &	$29635$ \\
$ $ &	$  $	&$ $ &	$ $ &	$ $ &$ $ &	$ $ &	$ $ &	$ $ &	$ $ 	&$ $ &	$ $ &	$  $&	$ $ &	$ $ &	$ $ &	$ $ &	$ $ &	$ $ &	$ $ &	$ $ &	$ $ 	&$  $&	$ $&	$ $ &	$ $ 	&$ $ &	$ $ &	$ $ &	$ $ &	$ $ &	$ $ 	&$232004$ &	$-71141$ &	$94864$ &	$23007$ \\
$  $	&$  $	&$  $&	$ $ &	$ $ &	$ $&	$ $ 	&$ $ &	$ $ 	&$  $&	$ $	&$ $&	$ $ &	$ $ &	$  $&	$ $ &	$  $&	$ $ 	&$ $ &	$ $ 	&$ $ 	&$ $ &	$ $ 	&$ $ &	$ $ &	$ $ &	$ $ 	&$ $ 	&$ $ &	$  $&	$ $ &	$ $ &	$ $ &	$490753$ &	$-256627$ &	$82484$ \\
$ $	&$ $ &	$ $ &	$ $ &	$ $ &$ $ &	$ $ &	$ $ 	&$ $ &	$ $ &	$ $&	$  $&	$ $ &	$  $&	$ $ 	&$ $ &	$  $	&$ $ &	$ $ 	&$ $ 	&$ $ &	$ $ &	$ $ &	$ $ 	&$ $ &$ $ &	$ $ &	$  $&	$ $ &	$ $ &	$  $	&$ $ 	&$ $ &	$ $ &	$872142$ &	$-77339$ \\
$ $	&$ $ &	$ $ &	$ $ &	$ $ &$ $ &	$ $ &	$ $ 	&$ $ &	$ $ &	$ $&	$  $&	$ $ &	$  $&	$ $ 	&$ $ &	$  $	&$ $ &	$ $ 	&$ $ 	&$ $ &	$ $ &	$ $ &	$ $ 	&$ $ &$ $ &	$ $ &	$  $&	$ $ &	$ $ &	$  $	&$ $ 	&$ $ &	$ $ &	$ $ &	$863104$ \\
\end{pmatrix}\times10^{-7}\qquad
$}
\end{center}
\end{sidewaystable}

\end{document}